\documentclass[12pt,preprint]{aastex}

\newcommand{\etal}{{\em et al.}}

\begin{document}


\title{The Next Generation of Photo-Detectors for Particle Astrophysics}


\author{Robert G. Wagner\altaffilmark{1}, Karen L. Byrum, Mayly Sanchez, Alexandre V. Vaniachine}
\affil{Argonne National Laboratory}

\author{Oswald Siegmund}
\affil{Space Sciences Laboratory, University of California, Berkeley}

\author{Nepomuk A. Otte}
\affil{University of California, Santa Cruz}

\author{Erik Ramberg, Jeter Hall}
\affil{Fermi National Accelerator Laboratory}

\and

\author{James Buckley}
\affil{Washington University, St. Louis}


\begin{abstract}
We advocate support of research aimed at developing alternatives to the
photomultiplier tube for photon detection in large astroparticle experiments such
as gamma-ray and neutrino astronomy, and direct dark matter detectors.
Specifically, we discuss the development of large area photocathode
microchannel plate photomultipliers and silicon photomultipliers.  Both
technologies have the potential to exhibit improved photon detection efficiency
compared to existing glass vacuum photomultiplier tubes.
\newpage
\end{abstract}


\section{Introduction}       \label{sec:intro}

High energy physics and astroparticle physics experiments require large area photon counting detectors with high
efficiency and low cost per channel cost. Currently, vacuum technology is the only way to achieve the large areas
while maintaining
high gains and photon-counting operation.  Even as the requirements of future experiments grow, the available
vendors of vacuum-based photon detectors are shrinking.  Companies that
manufacture vacuum tube based photon counting detectors are reducing their capabilities in favor of more
profitable semiconductor based technologies.  This threatens to raise the costs, extend the
planning required, and even compromise large scientific projects based on the detection of coherent light
phenomena, especially scintillation and Cherenkov light.  

The most common photon detection device in use for high energy and astroparticle physics is the photomultiplier tube
(PMT); first developed in the 1930s \citep{lub06}.  In addition to particle, nuclear and astrophysics
experiments, the technology is used today in a wide range of research fields such as space, medicine, biology, and
chemistry.  While the basic structure of PMTs remains unchanged from its inception, much development has taken
place with regard
to improved photocathode efficiency, precision timing characteristics, and multi-anode readout capability.
Depending on the details of the dynode structure, amplifications of $10^7$ can be achieved while
still retaining single photoelectron sensitivity.  Photocathode efficiencies presently are typically on the order of
20\%, although new photocathode technologies can achieve twice that value.  The
transit time for the generation of a signal is typically $<100$ns, with the transit time spread, and
thus, timing resolution for single photoelectrons, on the order of a nanosecond.  The cost of a phototube can be
as low as \$10/$\mbox{cm}^2$ of photocathode coverage, with optimized detectors costing up to 10 times that.
Alternative technologies to PMTs have arisen, but the superb performance and economics of
phototubes is such that major innovation for vacuum photomultipliers has proceeded slowly.

A promising new approach being pursued by several research groups is to combine modern semiconductor technology
with vacuum devices.  For example, new cathode
structures (in some cases semiconductor heterostructures engineered to promote efficient electron transport and other
features such as intrinsic high gain or fast timing) are being combined with new kinds of electron multipliers that
lend themselves to very large areas, and mass fabrication.  Designs for these electron multipliers range from
silicon and ceramic microchannel plates and semiconductor nanostructures to gas electron multipliers (GEMs) aimed at
dramatically reducing the cost per
unit area.  New techniques for direct deposition of cathode surfaces on silicon microchannels could result in
dramatic improvements.  The funding for all of these developments falls short of what is required to maintain
credible efforts at universities and national laboratories.  Development work at U.S. research institutions is lagging
compared to similar efforts based mainly in Europe.  A substantial change in the status quo is needed for new
technical approaches to be developed before further shrinkage of commercial technology occurs and for the U.S. to
remain competitive in the development of photon detection technology.

In this white paper we address two such technologies that hold the potential for replacing the PMT in many
astrophysical applications: large area microchannel plate photomultipliers (MCP) and Geiger-mode avalanche
photodiodes or silicon photomultipliers (SiPM).

\section{Physics Motivations}        \label{sec:motiv}

\subsection{Gamma-ray Astronomy}     \label{subsec:gamma-astro}

TeV Gamma-ray astronomy is one of the youngest sciences, having established its first firm detection, the Crab Nebula,
\citep{wee89} in 1988.  In the past 3 years, the number of TeV-band sources has increased by an order of
magnitude, including the remnants of supernova explosions, neutron stars, supermassive black holes, and
possibly associations of massive stars.  TeV gamma-ray astronomy has a broad science program which includes
cosmology and fundamental physics through studying of the nature of dark matter,
searching for a TeV component in Gamma-ray Bursts, determining the origin of cosmic rays, and studying jet
phenomena in supermassive black holes \citep{buc08}.
 
High-energy gamma rays can be observed from the ground by detecting secondary particles of the atmospheric
cascades initiated by the interaction of the gamma-ray with the atmosphere.  Imaging atmospheric Cherenkov
telescopes (IACTs) \citep{hin04,lor04,acc08} detect Cherenkov photons ($\lambda > 300$nm),
which are produced by electrons and positrons from the cascade.  The technique focuses Cherenkov photons onto
a finely pixelated camera typically composed of a few hundred phototubes and operating with an exposure of a few
nanoseconds.  The technique provides energy threshold triggering with energy resolution of $\sim15-20$\%.
The next generation of telescopes are attempting to reach sensitivities an order of magnitude greater than today's
instruments.   A major challenge is to scale observatories to this sensitivity while reducing the cost per
detector channel significantly from what would be the case by simply cloning
present day instrumentation to achieve an IACT coverage of $\sim 1 \mbox{km}^2$.  

\begin{figure}[t]
\begin{center}
\includegraphics[width=0.5\textwidth]{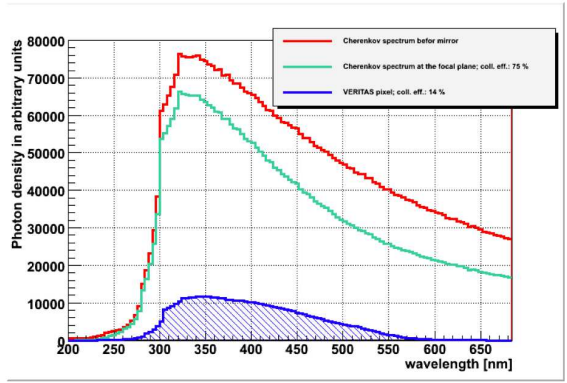}
\end{center}
\vspace*{-1.0cm}
\caption{\footnotesize{The Cherenkov spectrum emitted by particles of a 50 GeV gamma ray shower (red line).  The sharp
attenuation of the spectrum below 300nm is due to atmospheric absorption. The cyan line shows the Cherenkov
spectrum after two reflections on mirror surfaces and the blue hatched area shows the $\sim10$\% of the total Cherenkov
light arriving on the ground (here $\sim 1200$m) that is detected by standard bialkali photomultipliers typical of
those used in IACTs.} \label{fig:cspec}}
\vspace*{-0.25cm}
\end{figure}

The Cherenkov light pool from an atmospheric cascade consists of a large region (radius $\sim120$m) in which the
photon density is roughly constant and outside of which the photon density declines as a power law.  The
Cherenkov spectrum is shown in figure \ref{fig:cspec} for particles of a 50 GeV gamma ray shower.  Cherenkov light
arriving at the camera is detected by standard bialkali photomultipliers with UV glass entrance windows.
Typically only $\sim10$\% of the light is detected allowing for considerable improvement in sensitivity via enhanced
overall photon detection efficiency.
In addition to better photon detection efficiency, future IACT arrays seek to have finer angular resolution by
factors of 2-3 over present instruments.  The feasibility of incorporation of multi-anode phototubes into IACT
cameras has been demonstrated \citep{byr07}.  These would allow angular resolution on an event-by-event basis of
perhaps $0.05\degr$.  Cameras based on silicon photomultipliers may offer similar resolution with improved
photon detection efficiency and detector robustness.

\subsection{Cosmic-ray Astrophysics}

Over the past century, the study of cosmic radiation has had a significant impact on our understanding of nature.
The discovery of new fundamental particles in cosmic radiation, e.g. the positron and the muon, fundamentally
changed our knowledge of matter and led to the emergence of modern particle physics.   The continued study of
the cosmic radiation may yield significant information on supernova remnants, supermassive black holes,
and the nature of dark matter.  Additionally, cosmic ray outreach projects play an accessible role
in the education of young people on the methods of physical science.  Atmospheric fluorescence detectors, which are
critical in the highest energy cosmic-ray detectors such as HIRES and the Pierre Auger Observatory, need large area
photodetectors.

\subsection{Neutrinos}       \label{subsec:nu}

The confirmation and study of neutrino oscillation over the past two decades was a breakthrough
in our fundamental understanding of the nature of matter.  The observation of neutrinos from
supernova 1987a has improved our understanding of the supernova process while placing independent limits on neutrino
mass.   Neutrino detectors hold the promise of
answering some of our most fundamental questions about nature, but inexpensive, large area photodetectors are
needed to advance this key area of science over the next decade.

Two technologies stand out for the future of neutrino science: water Cherenkov detectors and liquid argon time
projection chambers \citep{bar07}.  Progress in water Cherenkov neutrino detection is aiming toward a million ton
detector.  The surface area surrounding the water volume is covered
with photon counting detectors, and at the megaton scale, such experiments face significant costs from covering
millions of square meters with photodetectors.

Liquid argon time projection chambers detect the neutrino produced lepton by drifting electrons liberated from
large argon volumes to wire grids.  Large area photodetectors enhance this technology by detecting the
scintillation light created during this process \citep{era06}.  The scintillation light is a much more accurate
timing process
for the neutrino interaction.  Since most neutrino sources, e.g. accelerators and supernovae,
have low duty-cycles, accurate timing will enable significant reduction of backgrounds, increasing the
speed and quality of scientific results from liquid argon neutrino detectors.

\subsection{Direct Dark Matter Detection}     \label{subsec:ddm}

A model for dark matter motivated by theoretical extension of the Standard Model of Particle Physics is the presence
of Weakly Interacting Massive Particles (WIMPs) produced in
the early universe and forming large extragalactic structures.  Direct detection of these very
weakly interacting particles may be accomplished using very large, radiopure detectors that can
detect nuclear recoils from elastic scattering of WIMPs.  These scatters would give only a few keV of recoil energy and
interaction rates could be as low as a few events per ton per year.
 
Detectors based on liquified noble elements have made promising breakthroughs during the past decade and may rapidly
increase the sensitivity to WIMP-nucleus recoils.  However, these technologies already
require specialized photon-counting devices.  Not only must the photo-detectors work at temperatures appropriate
for liquified xenon (165 K) or argon (87 K), but these devices must use less radioactive materials in their
composition; modern PMTs are already the most troublesome source of background for noble liquid dark matter detectors
that are in the construction and commissioning phase.  The next generation of liquid noble dark matter detectors
are expected to need many square meters of photosensitive area.
 
If the WIMP-nucleus scattering is not detectable by current experiments, a new large area photo-detector is
necessary for future noble liquid detectors.  As discussed in section \ref{subsec:mcp}, a possible solution
to large area photodetection is the Microchannel Plate photomultiplier.
These structures have a simple flat panel construction and, if they are made of low Z
materials, could be adequately radiopure.  The fast timing characteristics of these devices can contribute to
event localization.  Other solutions include hybrid photo-diodes, but both solutions are in the research
and development phase.  Future dark matter detectors based on liquid noble elements will be
significantly delayed during the next decade if the pace of research in large area photodetectors continues at
its current rate.

\section{Photodetection Technology in Astrophysics: Advantages and Limitations}     \label{sec:desc}

The advantages of PMTs in astronomy and astrophysics are high gain, fast signal formation, blue sensitivity, single
photon response, potential large sensitive area arrays, and the fact that it is a proven
technology with reasonable cost per channel or per photocathode coverage.  The limitations are the generally low
quantum efficiency, the difficulties in installation and support, e.g. the photocathode size is limited
and each tube is relatively bulky and delicate.  Moveover, high voltage needs to be supplied to each device often
requiring a massive overall cable burden.
Furthermore, PMTs under bias are damaged if accidently exposed to day light and are sensitive to magnetic fields.

\subsection{Large Area Photo-detectors Using Microchannel Plate Technology}     \label{subsec:mcp}

Progress in modern micro-electronics, materials science, and nano-technology provide an opportunity
to apply advanced microchannel plate technology to develop large-area economical photodetectors.   A
Microchannel Plate (MCP) is an array of miniaturized electron multipliers oriented parallel to each other.
An MCP typically consists of lead silicate glass processed to enhance secondary emission and to improve
detection efficiency for specific radiation: UV, soft x-ray, etc.  Commercial
devices are fabricated with channel diameters $\leq25\mu$m with the most advanced MCP having
$2\mu$m channels for high image detail and fast response time.
The fabrication of traditional glass MCPs is a costly multi-step process and large area MCPs are difficult to produce.

Anodization of aluminum can produce pores with diameters between 20 and 500nm, the use of anodized aluminum
oxide (AAO) membrane coated with conductive metal oxide as an MCP has been proposed \citep{dro06}.  This could be an
industrially economical process based on self assembly.  AAO feasibility studies have proceeded and initial tests
have shown promising results \citep{rou09}.
 
The use of Atomic Layer Deposition (ALD) for MCP's has been developed over the last several years, and a great
deal of work has been done understanding the basic issues and exploring the multi-dimensional parameter space
\footnote{see http://www.arradiance.com/}.   ALD provides the ability to deposit one molecular layer at a time and
create a layer of material with a specific work function. By using modern micro-electronics, materials science, and
nano-technology, one could create a complete photomultiplier inside of each microchannel pore.

Futhermore, new photocathode materials based on nano-technology also have the potential for comparable (or higher)
quantum efficiencies by optimizing the cathode surface morphology and dielectric constant, and thus, tailoring
the near-surface electric field such that it could significantly enhance photoelectron emission.  This approach
could offer an alternative to lowering the surface work-function of a conventional photocathode by fine
(and expensive) tuning of its already complex chemical composition.   Additional gains in overall detector
efficiency could be obtained by using nano-engineered photon-trapping surface geometries with reduced reflection
losses, thus benefiting from technologies now standard in the solar cell industry. 

\subsection{Silicon Photomultipliers}     \label{subsec:sipm}

Many of the disadvantages in everyday handling of PMTs do not apply to semiconductor photon detectors, for
example, vulnerability to mechanical stress or damage when exposed to strong light sources when biased. However,
the use of semiconductor photo-detectors in astroparticle physics has been limited mainly because devices are too
small and too costly.

\begin{figure}[t]
\begin{center}
\includegraphics[width=0.5\textwidth]{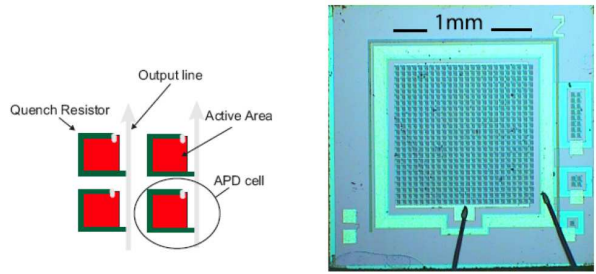}
\end{center}
\vspace*{-2.0cm}
\caption{\footnotesize{Schematic layout of  a SiPM; the figure shows four cells of a SiPM. Each cell is biased above
breakdown and connected via a quenching resistor to a common readout. Right: photograph of a SiPM produced by
MEPhI Pulsar.} \label{fig:sipm}}
\vspace*{-0.25cm}
\end{figure}

During the last decade, the possibility to use silicon photo-detectors has attracted increasing attention, mainly
due to the development of a novel detector concept: the Silicon Photomultiplier (SiPM) or the Geiger-mode APD.
The sensor, originally developed by several groups in Russia \citep{bon00}, is comprised of an array of avalanche
photo-diodes, typically $10^{2-4}$ per square millimeter
\footnote{see http://cerncourier.com/cws/article/cern/28805}.  Each photo-diode (or cell) is biased above
breakdown and operates in Geiger-mode, i.e. a photon is detected by a breakdown of a cell which generates a large
output pulse (typically several $10^{4-6}$ electrons). The breakdown is quenched by a polysilicon resistor
attached to each cell, which limits the current in the substrate below. The quenching resistor of each cell is
connected to a
common network so that the output signal of the SiPM is the summed signals of all cells (see figure \ref{fig:sipm}).
The SiPM offers many advantages such as low power, low voltage, compactness, excellent timing resolution, good
linearity, robustness, and insensitivity to magnetic fields. These characteristics make the SiPM an interesting
candidate for astrophysics experiments both on the ground and in space. 

SiPM applications to astroparticle physics experiments have been successfully demonstrated.
Most tests focused on its use in Cherenkov telescopes \citep{bil08,wag09} which is currently one
of the most demanding applications in terms of photon detection efficiency (PDE). One of these tests showed that
the performance of at least one type of SiPM is superior to the performance of a PMT with respect to PDE. The
potential of a PDE $>40$\% is one of the biggest attractions of SiPMs albeit a challenging goal.
Already, the PDE of some SiPMs outperforms that of the classical PMT with peak
efficiencies of about 30\% at 450nm. In the future it seems feasible that peak PDEs of 60\% and maybe higher
can be achieved.  The focus of future activities will be the development
of specially engineered entrance windows and avalanche structures which will enhance the sensitivity in the blue
to UV wavelength range. The design of new, blue sensitive structures depends on the ability to develop
processes which produce thin implantation layers at the surface to minimize the absorption of short wavelength
photons in insensitive areas just below the entrance window while avoiding a concomitant increase in noise rate
from the implantation layers.

The noise rate of SiPMs ranges between a few 100kHz and a few MHz per $\mbox{mm}^2$ sensor area at room
temperature.  Although a noise rate at the lower end, i.e. a few 100 kHz is sufficient for many applications in
astroparticle physics experiments, moderate cooling and temperature stabilization is required to compensate the
temperature dependent changes of SiPMs. For example, the gain changes by about $1\%/\degr C$.  With improved
processing,
for example, the replacement of the polysilicon resistor by a less temperature dependent material,
the temperature dependence can also be reduced.  Recent work at the Max Planck Institute/Munich, Germany has
demonstrated the feasibility of using a thermistor to automatically adjust the bias voltage to reduce gain variation
\citep{miy09}.  A separate group at the same institution is experimenting with
replacing the polysilicon quench resistor with a bulk integrated structure \citep{nin09}.  This would allow a
much improved geometric detection efficiency, i.e. eliminate the dead area the polysilicon produces.

Available SiPMs have sizes between $1\mbox{mm}^2$  and $25\mbox{mm}^2$. In astroparticle experiments photo-detector
size requirements typically range between 2cm to 40cm sized PMTs.  Whereas, for example, in Cherenkov telescopes SiPM
have about the right size, the application of SiPM in neutrino telescopes is limited.

\begin{figure}[t]
\begin{center}
\includegraphics[width=0.5\textwidth]{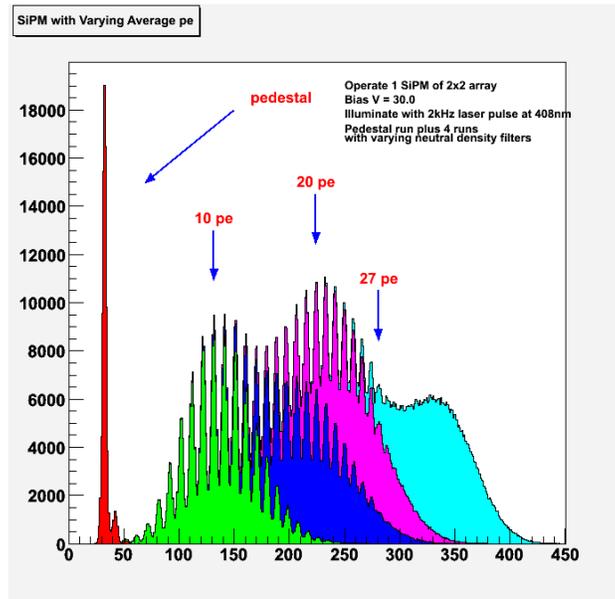}
\end{center}
\vspace*{-0.7cm}
\caption{\footnotesize{Pulse height distribution for a $5\times5\mbox{mm}^2$ SiPM exposed to a 2kHz laser pulse of
wavelength 408nm.  The distribution is the sum of a pedestal run plus 4 data runs with the average number
of incident photons varied using neutral density filters.  At least 27 individual photons per pulse firing
pixels in the photo-diode array are distinguished.  The limit on counting photons in this analog-to-digital
manner comes from noise and variance of the pulse height along with resolution limitation of the readout
electronics.}  \label{fig:sipm-sum}}
\vspace*{-0.25cm}
\end{figure}

One avenue of research for SiPMs is the possibility of incorporating them into an application specific
integrated circuit (ASIC).  SiPMs are inherently a digital device: each pixel is sensitive to interaction of a
single photon to which it produces a ``standard'' pulse with a rather small Gaussian variance about the mean.  The
ability to distinguish many incident photons is limited by the accumulation of noise, afterpulsing, and statistical
variance.  Even treating the output pulse as an analog sum of ``fired'' pixels and digitizing the signal, tens of
individual photons can be distinguished (c.f. figure \ref{fig:sipm-sum}).  Integration of the device into an ASIC
capable of actually sensing the number of pixels that absorbed a photon would take advantage of the digital nature
of the signal.  This would allow, for example, image reconstruction on IACTs by counting photons with a spatial
resolution on the camera plane of a few square millimeters.


\section{Conclusion}       \label{sec:conclu}

Traditional vacuum photomultiplier tubes are the ``workhorses'' of high energy gamma-ray, cosmic-ray,
and neutrino astrophysics.  While recent technical advances have improved the photon detection efficiency
of phototubes by as much as a factor of two, the number of suppliers of research quality PMTs is shrinking
and viable new alternatives have recently been developed.  We have briefly discussed two of these alternatives:
microchannel plate photomultipliers and silicon photomultipliers.  MCPs have the potential to be developed
into less costly large area photo-detectors for direct dark matter searches and neutrino astronomy.  SiPMs
with their higher photon detection efficiency, robustness (mechanical and upon exposure to high light levels),
and few square millimeter unit size appear to be a possible alternative for ground-based IACT camera
photo-detectors.  Both technologies are in need of significant further development to reach their potential.
Much work has been done by commercial manufacturers and research groups based mainly in Europe.  U.S.
universities and national laboratories could play a significant role in developing these technologies to
useful next generation photo-detectors for high energy astrophysics.  We advocate support for funding this
research and development and for supporting groups in the U.S. that are pursuing applied research with
MCPs and SiPMs.

\end{document}